\documentclass[lineno]{biometrikaArxiv}

\usepackage{amsmath,amssymb,graphicx,multirow, enumerate,color}

\usepackage{times}
\usepackage{bm}
\usepackage{natbib}

\usepackage[plain,noend]{algorithm2e}

\usepackage{chngcntr}
\usepackage{apptools}
\AtAppendix{\counterwithin{theorem}{section}}
\AtAppendix{\counterwithin{assumption}{section}}

\newcommand{\bx}{\textsf{x}}
\newcommand{\by}{\textsf{y}}
\newcommand{\bX}{\textsf{X}}

\newcommand{\bI}{\textsf{I}}
\newcommand{\bP}{\textsf{P}}
\newcommand{\bE}{\textsf{E}}
\newcommand{\bV}{\textsf{Var}}
\newcommand{\bCov}{\textsf{Cov}}

\DeclareSymbolFont{sfletters}{OML}{cmbrm}{m}{it}
\DeclareMathSymbol{\balpha}{\mathord}{sfletters}{"0B}
\DeclareMathSymbol{\bmu}{\mathord}{sfletters}{"16}

\makeatletter
\renewcommand{\algocf@captiontext}[2]{#1\algocf@typo. \AlCapFnt{}#2} % text of caption
\def\@algocf@capt@plain{top}
\renewcommand{\algocf@makecaption}[2]{%
  \addtolength{\hsize}{\algomargin}%
  \sbox\@tempboxa{\algocf@captiontext{#1}{#2}}%
  \ifdim\wd\@tempboxa >\hsize%     % if caption is longer than a line
    \hskip .5\algomargin%
    \parbox[t]{\hsize}{\algocf@captiontext{#1}{#2}}% then caption is not centered
  \else%
    \global\@minipagefalse%
    \hbox to\hsize{\box\@tempboxa}% else caption is centered
  \fi%
  \addtolength{\hsize}{-\algomargin}%
}
\makeatother

\def\Bka{{\it Biometrika}}

\begin{document}

\nolinenumbers  % use this command to turn off line numbers

%\jname{Biometrika}
%%% The year, volume, and number are determined on publication
%\jyear{2016}
%\jvol{103}
%\jnum{1}
%%% The \doi{...} and \accessdate commands are used by the production team
%%\doi{10.1093/biomet/asm023}
%\accessdate{Advance Access publication on 31 July 2016}
%
%%% These dates are usually set by the production team
%\received{April 2012}
%\revised{October 2015}

%% The left and right page headers are defined here:
\markboth{Hao Chen \and Dylan S. Small}{New tests for assessing covariate balance}

%% Here are the title, author names and addresses
\title{New multivariate tests for assessing covariate balance in matched observational studies}

\author{Hao Chen}
\affil{Department of Statistics, University of California, Davis \\ One Shields Ave., Davis, CA 95616 USA \email{hxchen@ucdavis.edu}}

\author{\and Dylan S. Small}
\affil{Department of Statistics, The Wharton School, University of Pennsylvania, \\ 3730 Walnut Street, Philadelphia, PA 19104 USA \email{dsmall@wharton.upenn.edu}}

\maketitle

\begin{abstract}
We propose new tests for assessing whether covariates in a treatment group and matched control group are balanced in observational studies.  The tests exhibit high power under a wide range of multivariate alternatives, some of which existing tests have little power for.  The asymptotic permutation null distributions of the proposed tests are studied and the $p$-values calculated through the asymptotic results work well in finite samples, facilitating the application of the test to large data sets.  The tests are illustrated in a study of the effect of smoking on blood lead levels.  The proposed tests are implemented in an \texttt{R} package \texttt{BalanceCheck}.
\end{abstract}


\section{Introduction}
In testing the effect of treatments, random assignment is desirable because it tends to make the treated group and the control group similar in both observed and unobserved covariates.  However, in many applications, for practical and ethical reasons, we may not be able to assign subjects randomly, such as in studying the effect of smoking.  Rather, we must conduct an observational study.  In a matched observational study, we hope to measure all the confounders and then construct a treated (e.g., smoker) group and a matched control (e.g., not smoker) group that are similar with respect to these confounders.  If there is only one confounder, it may be feasible to match treated units exactly to control units, but there are often many confounders in which case matching exactly is infeasible.  Typically, we cannot find exact matches when there are many confounders, but we could still hope that the distribution of the covariates is similar in the treated and matched control group -- this is what a randomized experiment hopes to achieve, not to create identical treated and control groups.

A key question for a matched observational study is, has the matching created a treated group and a matched control group that have similar distributions?  If the answer is yes, then any substantial differences in outcomes between the two groups can be attributed to the treatment while if the answer is no, then a difference in outcomes between the two groups may just be due to pretreatment differences between the groups.  A commonly used approach for comparing the treated and matched control groups is to compare the distribution of the measured confounders of the two groups (e.g., gender, age, height, weight, profession, etc.) one by one, for example by conducting two sample $t$-tests.  A problem with this univariate approach is that the joint distributions of the covariates might be quite different between the two groups even though their marginal distributions are similar.  While we may still want to keep the individual tests for covariates separately as these tests are usually powerful in detecting marginal differences, we would like to have a more holistic test to take care of the joint distribution of the covariates.

 \cite{rosenbaum1985constructing} proposed to use the propensity score, the probability of a subject being assigned to a particular treatment given a set of covariates, to check covariate balance.  By summarizing all observed covariates into one value, it greatly simplifies the process and has been applied in many studies \citep{berk1985does, hoffer1985achievement, myers1987time, fiebach1990outcomes, czajka1992projecting, stone1995propensity, lieberman1996association}.
%can be done on a large number of covariates.  Hence, starting from its launch, it has been used in many studies to reduce bias and control confounding factors
However, to estimate the propensity score, one needs to model the distribution of the treatment indicator variable given the observed covariates.  So choosing a good model is important for this approach. % The most widely used model in practice is the ordinary logistic regression \citep{d1998tutorial}.
 More recently, some less model-based approaches have been proposed and used in several studies.    \cite{hansen2008covariate} advocated a randomization inference cousin of Hotelling's $T^2$ statistic which is called the method of combined differences.  A reviewer suggested seemingly unrelated regression as a related approach.  \cite{hansen2008covariate} focus on the goal of evaluating
  whether the matched pairs in a matched observational study resemble a matched pair randomized experiment (with respect to the observed covariates).  In this paper, we focus on the goal of evaluating whether the treated and matched control group in a matched observational study resemble a balanced simply randomized experiment; for discussion of settings where this goal is useful, see \cite[\S 1.2]{rosenbaum2013} and \cite[Ch. 15]{imbens2015}.  \cite{gagnon2016} propose using a classifier (e.g., logistic regression,
random forests) to distinguish treated units from control units and then use permutation inference to test whether the classifier is in fact able to distinguish treated units from control units more accurately than would be expected by chance.
  \cite{heller2010using} proposed to use the crossmatch test of \cite{rosenbaum2005exact} to test covariate balance through forming a non-bipartite graph among the treatment and control subjects.  Distributions that are very different between the treated and control groups will exhibit few cross-matches (matches between treated and control subjects).
 %proposed to test through a bipartite graph between the treatment and control subjects.  Distributions that are very different will exhibit few cross-matches.

 %advocate the use of Fisher's randomization inference for balance checking.  % These tests are useful in diagnosing whether the two groups are similar and achieved success in XXX.

When the number of covariates is large, the behavior of ``closeness" could be counterintuitive due to the curse of dimensionality.  The existing methods more or less use the implicit assumption that ``closer'' observations are more likely to be from the same distribution.  This assumption may not hold when the dimension of the observation (number of covariates) is high \citep{chen2017new}.
In this paper, we propose new tests to address this problem in covariate balance assessment.  The proposed tests still use similarity information among the observations, while the test statistics are designed so that more alternative scenarios could be covered under typical sample sizes, as well as taking into account the effect of matching. % Due to the existence of the matching step, the behavior of the two samples is different from two independent samples.  The proposed tests also take this into account and exclude
The proposed tests are implemented in an \texttt{R} package \texttt{BalanceCheck}.

The rest of the paper is organized as follows.  Section \ref{sec:review} reviews existing two-sample tests based on similarity information.  Section \ref{sec:test} discusses the proposed tests and their asymptotic properties.  The power of the proposed tests is evaluated and compared to existing tests in Section \ref{sec:power}.   The proposed tests are applied to an observational study on the effect of smoking on blood lead levels in Section \ref{sec:application}.  The paper concludes with discussion in Section \ref{sec:conclusion}.

% The proposed tests rely on how similar the subjects are in terms of the covariates, and thus have few distributional assumptions on the joint distribution of the covariates.  Existing tests based on similarity information are reviewed in Section \ref{sec:review}.  The proposed tests, by taking into account a wider range of alternatives, as well as the nature of matching, exhibit higher or comparable power than the existing tests.
% The details of the proposed tests, as well as their asymptotic properties, are provided in Section \ref{sec:test}.  In Section \ref{sec:power}, the power of the proposed tests is evaluated and compared to existing tests.  In Section \ref{sec:application}, the tests are applied to an observational study on the effect of smoking on blood lead levels.  The paper concludes with discussion in Section \ref{sec:conclusion}.

\section{A brief review of two-sample tests based on similarity information}
\label{sec:review}

The setup for general two independent sample testing is as follows: There are two independent samples $\{\bx_1,\dots,\bx_n\}$ and $\{\by_1,\dots,\by_m\}$, where the units in each sample are independent and identically distributed according to distributions $F_\bx$ and $F_\by$, respectively.  The setup for covariate balance assessment is slightly different as the two samples are no longer independent due to the matching step. % . , while some ideas could still be adopted.
We here review the existing two independent sample tests based on similarity information as the proposed tests build up on them.

% When the dimension of the data is low, parametric methods can often be employed.  However, the power of parametric methods decreases quickly as the dimension increases unless strong assumptions are made to avoid the estimation of large number of parameters.  We here review the non-parametric approach based on the similarity information among the observations.

\cite{friedman1979multivariate} developed the first practical two-sample test for multivariate data based on similarity information.  The test considers the minimal spanning tree (MST) constructed from the pairwise distances among the pooled observations from the two samples, which is the spanning tree connecting all observations from the two samples that minimizes the sum of distances of edges in the tree.  The test statistic is the number of edges that connect nodes (observations) from different samples and the null hypothesis of equal distribution is rejected if this number is significantly {\it{less}} than its null expectation.  The idea behind the test is that, if the two samples are from the same distribution, the two samples are well mixed and we expect a relatively large number of edges that connect subjects from different samples.  So if the number of between-sample edges is small, this suggests the two samples are from different distributions.  We call this test the \emph{edge-count test} for easy reference.

% The edge-count test can be applied to other similarity graphs as well.  \cite{friedman1979multivariate} also tried on $k$-MST, which is the union of the 1st, \dots, $k$th MSTs, where the $i$th MST is a spanning tree connecting all observations that minimizes the sum of distances across edges subject to the constraint that the spanning tree does not contain any edge in the 1st, \dots, $i$-1th MST(s).  They showed that the edge-count test on a 3-MST is usually more powerful than that on a 1-MST.  \cite{schilling1986multivariate} and \cite{henze1988multivariate} used $k$-nearest neighbor graphs where each observation is connected to its $k$ closest neighbors.

% \cite{chen2017new} considered testing for $H_0: F_\bx = F_\by$ against a general alternative $H_A: F_\bx\neq F_\by.$  Their test is based on a similarity graph constructed on the pooled observations, such as a minimum spanning tree (MST) where all observations are connected and the sum of distances of all connecting edges are minimized.

% Two-sample tests based on similarity graphs were first proposed by \cite{friedman1979multivariate} where the MST is used.

The edge-count test was later extended to other similarity graphs, such as nearest neighbor (NN) graphs in \cite{schilling1986multivariate} and \cite{henze1988multivariate}, and a non-bipartite matching graph in \cite{rosenbaum2005exact} (crossmatch).  Extensions of the edge-count test have been applied to assess covariate balance in several observational studies \citep{heller2010using, mchugh2013hospitals}.

\cite{maa1996reducing} showed that the edge-count test based on the MST constructed on Euclidean distance is consistent against all alternatives for multivariate data. % In all these tests, the test statistic is the number of edges in the similarity graph that connect observations between the two samples and the null hypothesis is rejected if this count is significantly \emph{smaller} than its expectation under permutation null distribution.  The rationale is that, if the two samples are from the same distribution, the observations are well mixed and the number of between-sample edges would be large.  By rejecting scenarios with small numbers of between-sample edges, the tests are hoped to be powerful under general alternatives.
Nevertheless, in a recent study, \cite{chen2017new} found that the edge-count test has low or even no power under some common alternatives in practical sample sizes.  For example, when the two distributions are $F_\bx = \mathcal{N}(\mathbf{0}, I_d)$ and $F_\by=\mathcal{N}(\bmu,\sigma^2 I_d)$, $\|\bmu\|_2=1, \sigma=1.1, d=100, n=m=1,000$, the number of between-sample edges in the MST constructed on the pooled 2,000 observations is close to its null expectation in typical runs.  In contrast, if $\sigma=1$ and all other parameters are the same, the number of between-sample edges is significantly smaller than its null expectation in typical runs.  This is a weird phenomenon {\textendash}  the MST test has more power when the distributions are less different.  The underlying reason is the curse of dimensionality: The volume of the $d$-dimensional ball grows exponentially in dimension, so the observations from the sample with a slightly higher variance tends to find the observations from the sample with a slightly smaller variance to be closer, resulting in the large number of between-sample edges.

\cite{chen2017new} solved this problem for the two independent sample testing setting by proposing a modified test statistic based on the following fact: In the first scenario above where the edge-count tests work, which we call `Scenario (a)', the number of within-sample edges for both samples are significantly \emph{larger} than their null expectations.  In the second scenario above where the edge-count test does not work, which we call `Scenario (b)', the number of within-sample edges for the sample with slightly smaller variance is significantly \emph{larger} than its null expectation, while the number of within-sample edges for the sample with slightly larger variance is significantly \emph{smaller} than its null expectation.  The test proposed in \cite{chen2017new} aggregates deviations in both directions of the within-sample edges for the two samples from their null expectations, and thus works for more alternatives than the tests that only rely on between-sample edges.  We call this test the \emph{generalized edge-count test} for easy reference. % \cite{chen2017new} showed it is indeed the case.

% The edge-count test has been applied to assess covariate balance in several studies \citep{heller2010using, mchugh2013hospitals}.
Although the generalized-edge count gains power in the two independent sample testing setting, it can lead to wrong conclusions under the setting of covariate balance assessment in observational studies because the null includes settings with more balance than expected with independent samples from two identical distributions.  Consider the following toy setting: There are 30 covariates of interest, and these 30 covariates follow a standard multivariate Gaussian distribution for subjects both in the treatment group and the control group.  100 subjects are randomly selected from the treatment group and 300 from the control group.  We first compare the covariate balance of the two groups directly; then, we select 100 subjects from the control group that match the 100 subjects in the treatment group the best through the \texttt{Match()} function in \texttt{R} package \texttt{Matching}.  Since here the distributions of the covariates are the same for the treated subjects and controls, we would hope the test on covariate balance would usually not to be rejected after matching.  We repeat the simulation run for 1000 times and record the percentages of trials that the generalized edge-count test rejects the null hypothesis of covariate balance at 0.05 level (Table \ref{tab:CF}).

\begin{table}[!htp]
\caption{The proportion of trials (out of 1000) that the generalized edge-count test rejects covariate balance at the significance level 0.05.} \label{tab:CF}
\begin{center}
\begin{tabular}{|c|c|}
\hline
Before Matching & After Matching
\\ \hline \hline
0.043 & 0.549 \\ \hline
\end{tabular}
\end{center}
\end{table}

From Table \ref{tab:CF}, we see that the generalized edge-count test is behaving properly before matching in that its size is about equal to the desired .05.  Intuitively, after matching, the covariates should be even more balanced than before and the size should be at most .05.  However, after matching, the generalized edge-count test has a very high rejection proportion.  This is counterintuitive, but the reason lies in how the test was constructed.
% This shed lights on having a powerful test under a wider range of alternatives in comparing the covariates of the two groups in observational studies.  If the two groups, treated group and control group, are randomly selected from their own populations, the test in \cite{chen2017new} is directly applicable.  However, in most observational studies, in order to have two groups that are similar in covariates, a matching is usually done first, i.e., researchers search over the control group to find the most similar controls to be included in the study.  In this case, the test in \cite{chen2017new} cannot be directly applied.
Consider the scenario that the matching is done very well.  If we construct a MST on the observations, many edges would be connections between the two groups, i.e., both numbers of within-sample edges would be small.  The test in \cite{chen2017new} rejects in this scenario as these numbers significantly deviate from their null expectations.
% Is it okay to just leave the next two sentences out for keeping things concise?  Matching is typically not perfect or close to perfect although the balance is considerably
% better than that achieved by just having two identical independent samples.
% This issue does not exist for testing two independent samples because close to perfect matching for two randomly selected samples happens with extremely low probability.  However, this is common in matching, especially when the matching is done well.

In the following, we propose tests that have substantial power under both scenarios (a) and (b), thus gaining power over existing similarity tests of covariate balance assessment, while tending not to reject in the scenario that the matching balances the covariates well.

% Therefore, we need to have a new test statistic for covariate balance assessment when a matching is done.  We want the test to inherit the desirable property of the test in \cite{chen2017new} while not to reject the scenarios that the matching is done well.

\section{New tests}
\label{sec:test}
% $\max(R_1,R_2)$, $\max(D_1, D_2)$.
% $D_1$: the number $k$-NN the treatment group find within themselves.
% $D_2$: the number of $k$-NN the control group find within themselves.
Let $n$ be the number of treated subjects (subjects exposed to a treatment, such as smoking) and $\by_i, \ i=1, \dots, n$, be the vector of observed covariates for subject $i$.  There are more than $n$ controls and among them $n$ subjects are matched to the treated subjects according to some matching algorithm.  Let $\by_{n+1},\dots,\by_{2n}$ be the observed covariate vectors of the matched controls.  Let $N=2n$ be the total sample size.  We test whether the two sets of covariates are similar through testing whether the group identities, treatment and control, ares exchangeable.  In other words, we work under the permutation null distribution that places $1/\binom{N}{n}$ probability on each of the $\binom{N}{n}$ choices of $n$ out of $N$ observations as the treatment group.  By considering this null distribution, we are considering whether the covariate balance is comparable to that of a randomized experiment, a useful and recognizable benchmark for appraising a matched comparison\citep{heller2010using}.  When there is no further specification, we denote by $\bP, \bE, \bV$, probability, expectation, and variance,
respectively, under the permutation null distribution.

We propose two new tests with one based on the nearest neighbor (NN) information of the subjects (Section \ref{sec:crossNN}) and the other based on the MST constructed on the pooled subjects (Section \ref{sec:crossMST}).  We call the tests ``CrossNN" and ``CrossMST", respectively.

\subsection{CrossNN}
\label{sec:crossNN}

We pool the $N=2n$ observations together and for each observation, finds its NN among the pooled $N$ observations.  Let $a_{ij} = 1$ if $\by_j$ is the NN of $\by_i$ and 0 if otherwise.  Let $D_{11}$ be the number of observations in treatment group whose NNs are from the treatment group as well, and $D_{22}$ be the number of observations in controls whose NNs are from the controls.  For any event $x$, let $\bI_x$ be the indicator function that equals 1 if event $x$ occurs and 0 if otherwise.  Then, $D_{11}$ and $D_{22}$ can be expressed as
\begin{align}
D_{11} &= \sum_{i=1}^n \sum_{j=1}^{n} a_{ij} = \sum_{i=1}^{N}\sum_{j=1}^{N} a_{ij}\,\bI_{i\in[1,n]}\,\bI_{j\in[1,n]}, \\
D_{22} &= \sum_{i=n+1}^N \sum_{j=n+1}^{N} a_{ij} = \sum_{i=1}^{N}\sum_{j=1}^{N} a_{ij}\,\bI_{i\in[n+1,N]}\,\bI_{j\in[n+1,N]}.
\end{align}

We propose to use $D_M=\max(D_{11}, D_{22})$ as the test statistic.  We reject the null of balanced covariates if $D_M$ is significantly \emph{larger} than its expectation under the permutation null distribution.  The statistic defined in this way has power for scenarios (a) and (b) because at least one of $D_{11}$ and $D_{22}$ is large under
the alternatives of either scenario.  On the other hand, under the perfect matching scenario, both $D_{11}$ and $D_{22}$ are small and the null of balanced covariates is not rejected.

The next question is how large $D_M$ needs to be to reject the null of covariate balance.  When $n$ is small, we can permute the group label directly to obtain the permutation distribution of $D_M$ and compute the permutation $p$-value.  However, when $n$ is large, direct permutation would be too time consuming.  We thus seek to find an analytic approach to do so.

\begin{figure}[!htp]
\begin{center}
\includegraphics[width=.40\textwidth]{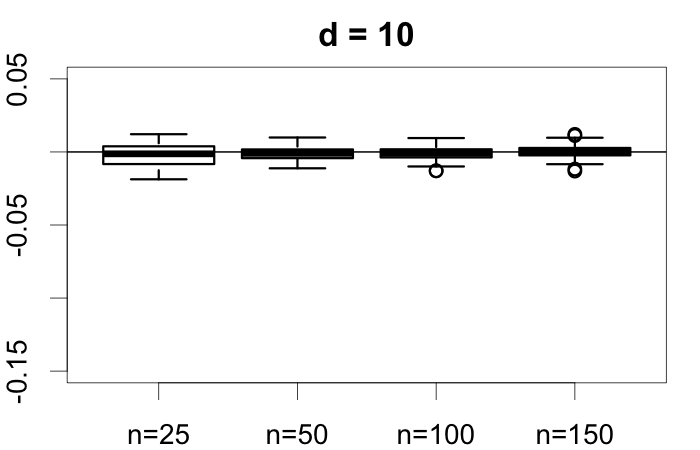} \quad
\includegraphics[width=.40\textwidth]{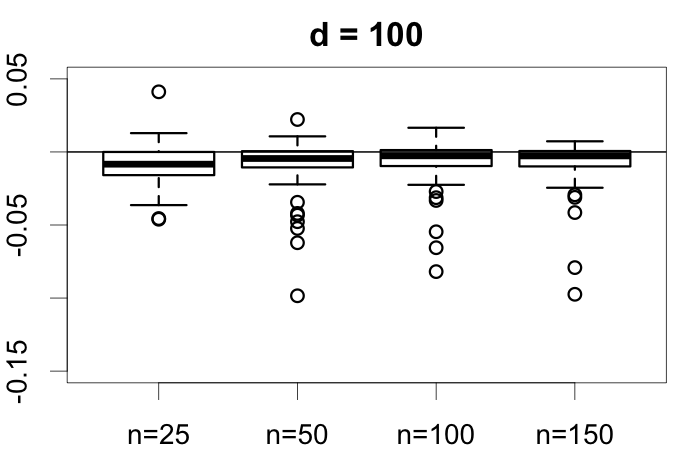}
\end{center}
\caption{Boxplots of the differences between the asymptotic $p$-value and the $p$-value calculated from 10,000 permutations (asymptotic $p$-value minus permutation $p$-value).}\label{fig:pvalue}
\end{figure}

In the following, we first derive exact analytic expressions for the expectations and variances of $D_{11}$ and $D_{22}$, as well as their covariance.  We then study the joint distribution of $D_{11}$ and $D_{22}$ asymptotically, which equips us to determine the asymptotic $p$-value for the test.  The asymptotic $p$-value is then compared to the $p$-value calculated through 10,000 permutations directly.  It turned out that the asymptotic $p$-value is reasonably accurate for sample sizes in the hundreds (see Figure \ref{fig:pvalue}).

% The main results are given in Lemma \ref{lemma:EV} and Theorem \ref{thm:asym}.
% For simplicity, in the following, when the summation range is not explicitly written out in the following, it is summing from 1 to $N$.

\begin{lemma}\label{lemma:EV}
The expectations and variances of $D_{11}$ and $D_{22}$, as well as their covariance are as follows:
\begin{align}
& \bE(D_{11}) = \bE(D_{22}) = \frac{N(N-2)}{4(N-1)}, \\
& \bV(D_{11}) = \bV(D_{22}) \nonumber \\
& \hspace{7mm} = \frac{1}{16}\left(\frac{N^2}{N-1}-\frac{N^2}{(N-1)^2} + 2C_1 \frac{N(N-2)}{(N-1)(N-3)} +2C_2\frac{N(N-4)}{(N-1)(N-3)} \right),  \\
& \bCov(D_{11},D_{22}) = \frac{1}{16}\left(\frac{N^2}{N-1}-\frac{N^2}{(N-1)^2} + (2C_1-2C_2) \frac{N(N-2)}{(N-1)(N-3)} \right),
\end{align}
where $C_1=\sum_{i=1}^{N-1} \sum_{j=i+1}^N a_{ij} a_{ji}$ is the number of pairs of observations that are mutual NNs and $C_2 = \sum_{i=1}^N \sum_{j=1}^{N-1} \sum_{l=j+1}^N a_{ji} a_{li}$ is
the number of pairs of observations sharing their NNs.
\end{lemma}
The lemma is proved through combinatorial analysis and the details are in supplemental materials.

\begin{remark}
According to Propositions 1 and 2 in \citep{henze1988multivariate}, when $N\rightarrow \infty$, we have $2C_1/N$ and $2C_2/N$ converge to constants that only depend on the dimension and the norm used to calculate the distance.  We denote the two limits by $q_1$ and $q_2$, respectively.
\end{remark}

Let $$Z_{D,1} = \frac{D_{11}-\bE(D_{11})}{\sqrt{\bV(D_{11})}}, \quad Z_{D,2} = \frac{D_{22}-\bE(D_{22})}{\sqrt{\bV(D_{22})}}.$$
Then the test statistic $Z_D = \max(Z_{D,1},Z_{D,2})$ is equivalent to $D_M$.  The following theorem states the asymptotic joint distribution of $Z_{D,1}$ and $Z_{D,2}$.

\begin{theorem}\label{thm:asym}
Under permutation null distribution, as $N\rightarrow\infty$, $(Z_{D,1},Z_{D,2})^\prime$
% \begin{equation}\label{eq:vector}
% % \left(\frac{D_{12}-\bE(D_{12})}{\sqrt{\bV(D_{12})}}, \frac{D_{21}-\bE(D_{21})}{\sqrt{\bV(D_{21})}} \right)^\prime
% \left(\frac{D_{12}-\mu}{\sigma}, \frac{D_{21}-\mu}{\sigma} \right)^\prime
% \end{equation}
converges to a bivariate Gaussian distribution with mean $\mathbf{0}$ and covariance matrix $\left[\begin{array}{cc} 1 & \rho_D \\ \rho_D & 1 \end{array}\right]$, where $\rho_D = (1+q_1-q_2)/(1+q_1+q_2)$.
\end{theorem}
The proof of this Theorem extends the methods in \cite{chen2013graph} and \cite{chen2017new}.  The complete proof is in supplemental materials.

Based on this Theorem, we can easily calculate the asymptotic $p$-value of $Z_D$.  That is,
\begin{align*}
P(Z_D\geq z) & = 1-P(Z_D<z) = 1-P(Z_{D,1}<z, Z_{D,2}<z),
\end{align*}
and $P(Z_{D,1}<z, Z_{D,2}<z)$ can be calculated from function \texttt{pmvnorm()} in the \texttt{R} package \texttt{mvtnorm}.  In practice, we use the finite sample version of the covariance between $Z_{D,1}$ and $Z_{D,2}$, that is, we use $$\rho_{D,N}=\frac{N(N-3)/(N-1) + 2C_1 - 2C_2}{N(N-3)/(N-1) + 2C_1 + 2C_2 (N-4)/(N-2)}$$ rather than $\rho_D$ when calculating $P(Z_{D,1}<z, Z_{D,2}<z)$.  It is easy to see that $\lim_{N\rightarrow\infty}\rho_{D,N} = \rho_D$.

% \subsubsection{Accuracy of asymptotic results on finite samples}
% \label{sec:accuracy}

% We next check how the asymptotic $p$-value approximation works for finite $n$.  The simulation setup is as follows: Two samples, each with $n$ observations, are generated independently from $d$-dimensional Gaussian distribution.  The test statistic $Z_m$ and its asymptotic $p$-value are calculated as described in Section \ref{sec:permutation}.  We run 500 simulation runs and calculate the fraction of runs that the asymptotic $p$-value is less than 0.05.  We then repeat this for 20 times.  The mean of the fractions over the 20 repeats, as well as twice the standard deviation of the mean, is plotted in Figure \ref{fig:pvalue}.

We compared the asymptotic $p$-values to $p$-values calculated from 10,000 permutations.  Figure \ref{fig:pvalue} plots the boxplots of the differences between the two $p$-values (asymptotic $p$-value minus permutation $p$-value) over 100 simulation runs under difference choices of $n$ and under different dimensions.  We see that when the dimension is low ($d=10$), the asymptotic $p$-value is already quite accurate for very small sample sizes.  The accuracy for very small sample sizes is not as good when the dimension becomes high ($d=100$), but it is reasonably accurate once the sample size is at least 100.

% becomes worse when the dimension becomes high ($d=100$), while it is still reasonable accurate when the sample size is in hundreds.

% We see from the figure that the asymptotic $p$-value becomes quite accurate when $n\geq 100$ for both low and high dimensions.  For $n<100$, the asymptotic $p$-value is slightly conservative.

% \subsection{Consistency of the test}
% \begin{theorem}
% For testing two independent samples with equal sizes, the test based on $Z_m$ is consistent against all alternatives.
% \end{theorem}

\subsection{CrossMST}
\label{sec:crossMST}
We pool the $n$ treated subjects and $n$ matched controls together and construct a MST on the pooled observations.  Let $R_1$ be the number of edges in the MST that connect subjects from the treatment group and $R_2$ be the number of edges that connect subjects from the control group.  We propose to use $R_M = \max(R_1,R_2)$ as the test statistic.  We reject the null of balanced covariates if $R_M$ is significantly \emph{larger} than its expectation under the permutation null distribution.

The statistic defined in this way has power for scenarios (a) and (b) because at least one of $R_1$ and $R_2$ is large under
the alternatives of each scenario.  On the other hand, under the perfect matching scenario, both $R_1$ and $R_2$ are small and the null of balanced covariates is not rejected.  Similar to $D_M$, we also studied the permutation null distribution of $R_M$ and the $p$-value approximation based on asymptotic results is also reasonable accurate for sample sizes in the hundreds.  The details for these are in Appendix \ref{sec:permutation_crossMST}.

\section{Power assessment}
\label{sec:power}

We now evaluate the power of the proposed tests, CrossNN and CrossMST, in comparison to three existing tests: Hotelling's $T^2$ test, the method of  combined differences proposed in \cite{hansen2008covariate}, and the crossmatch test proposed in \cite{rosenbaum2005exact}.  We follow a similar simulation setting to that in \cite{franklin2014metrics}.

The detailed setup is as follows: There are 6 observed covariates that are independent standard normal distributed ($X_1, X_2, X_3, X_4, X_5, X_6$).  The exposure/treatment, $T$, depends on these 6 covariates as well as the second order of four of them ($X_7=X_1^2, X_8=X_2^2, X_9=X_3^2, X_{10}=X_4^2$).  In particular, $T$ was simulated as a binary variable via the logistic model logit\{P($T_i=1$)\}$=\alpha_0 + \balpha \bX_i$, where $\bX_i$ is the covariate vector (including all 10 covariates) for subject $i$ and $\balpha=(\alpha_1,\dots,\alpha_{10})$.  The parameters in $\balpha$ define the log-odds ratios between covariates and exposure in the pre-matched dataset, and higher absolute values generally indicate more imbalance.  We generated 1,000 subjects and determined whether each of them is exposed or not.  Then, we matched control subjects ($T=0$) to exposed subjects ($T=1$) through their propensity scores on $X_1,\dots,X_6$, the main effects of the covariates.

To make the comparison simple, we let $\alpha_1=\dots=\alpha_6 = a$ and $\alpha_7=\dots=\alpha_{10}=b$.  We considered three scenarios: (i) there are only main effects ($a=0.4, b=0$); (ii) there are both main and second order effects ($a=b=0.4$); and (iii) there are only second order effects ($a=0, b=0.4$).  Since the propensity score is calculated based on the main effects, we would expect the matching could balance the covariates when there are only some main effects.  However, if there are second order effects, the matching would not be able to address them.

We also considered correlated data where $X_1, X_2, X_3, X_4, X_5, X_6$ are jointly normal $\mathcal{N}_6(0, \Sigma)$ with the $(i,j)$th element of $\Sigma$ being $0.1^{|i-j|}$.  The other setups are the same and the parameters are set to be $a=0.4$ and $b=0.4$.  This is noted by Scenario (iv).

\begin{figure}[!htp]
\centering
\includegraphics[width=.48\textwidth]{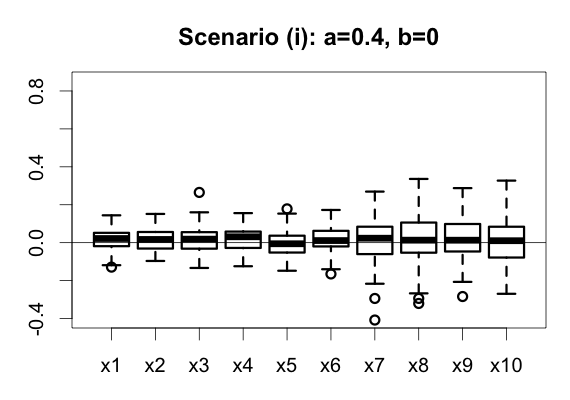}
\includegraphics[width=.48\textwidth]{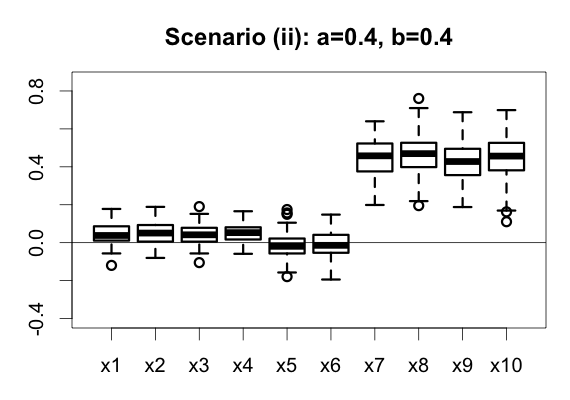}
\includegraphics[width=.48\textwidth]{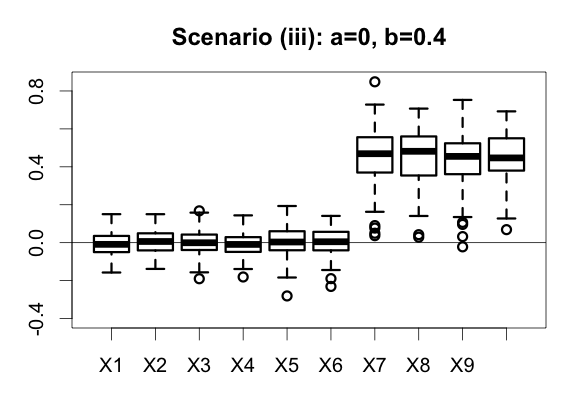}
\includegraphics[width=.48\textwidth]{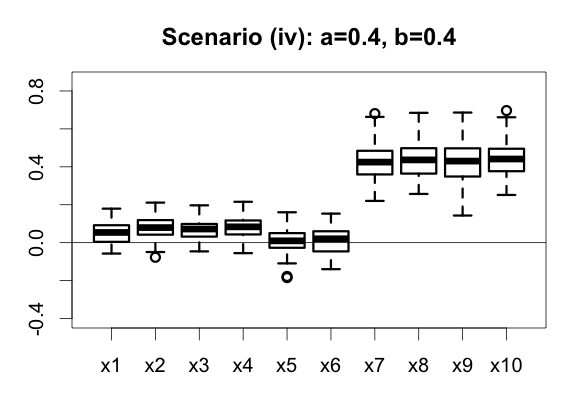}

\caption{Boxplots of standardized differences between the treatment group and matched control group over 100 trials for each scenario. \label{fig:SD}} %Standardized difference $=(\bar{x}_1-\bar{x}_2)/\sqrt{(s_1^2+s_2^2)/2}$, where $\bar{x}_m$ and $s_m^2$ are sample mean and variance for smokers $(m=1)$ and non-smokers $(m=2)$.}
\end{figure}

For each scenario, we simulated 100 data sets and calculated the standardized difference between the treatment group and matched control group for each of the ten covariates for each data set.  The standardized difference is defined as $=(\bar{x}_1-\bar{x}_2)/\sqrt{(s_1^2+s_2^2)/2}$, where $\bar{x}_m$ and $s_m^2$ are sample mean and variance for treated subjects ($m=1$) and matched controls ($m=2$).  Figure \ref{fig:SD} plots the boxplots of standardized differences of the 100 data sets for each of the ten covariates.  We see that the main effects ($X_1,\dots,X_6$) are well balanced in all scenarios.  So the matching over propensity scores is doing a good job for covariates included in the model.   However, second order effects ($X_7,\dots,X_{10}$) are very unbalanced except for scenario (i).  This is expected as they are not included in the model for calculating the propensity score and only scenario (i) does not have second order effects from the data generating step.

% We considered the scenarios that there are only main effects ($\alpha_7=\alpha_8=\alpha_9=\alpha_{10}=0$), there are both main and second order effects, and there are only second order effects.  To make the comparison simple, we let $\alpha_1=\dots=\alpha_6 = a$ and $\alpha_7=\dots=\alpha_{10}=b$.

Whether one wants to balance a covariate depends on whether the covariate contributes to a prognostic score for one or more outcome of interest \citep{hansen2008covariate}.  In the context of our simulation study, if a second order effect contributes to a prognostic score for one or more outcomes of interest, we would like to balance it, otherwise not.  Suppose that the second order effects $X_7=X_1^2, X_8=X_2^2, X_9=X_3^2, X_{10}=X_4^2$ all contribute to a prognostic score.  Then, we want a test that will not reject covariate balance for scenario (i), while it would reject covariate balance for scenarios (ii), (iii) and (iv).

Table \ref{tab:power} shows the proportion of trials, out of 100, that the test reject covariate balance at the significance level 0.05 for the three scenarios.  In the table, the Hotelling's $T^2$ test is denoted by $T^2$ and the method of combined differences is denoted by CD.  They are applied to compare the 6 main effects ($T^2$ and CD in the table) and to compare 6 main effects and all 15 of their two-way interactions ($T^2 \times$ and CD $\times$ in the table).  The crossmatch test, the CrossNN test, and the CrossMST test are denoted by CM, NN, and MST in the table.  They are applied to the 6 main effects only.  Two distances are considered: the rank-based Mahalanobis distance (CM$_1$, NN$_1$, and MST$_1$ in the table), and the $L_2$ distance (CM$_2$, NN$_2$, and MST$_2$ in the table).

From the table, we see that all tests are doing a good job for scenario (i) that the proportions of rejections are all controlled by the significance level.  For scenarios (ii), (iii) and (iv), we want a high rejection proportion.  We see that the Hotelling's $T^2$ and the method of combined differences do not have power in these two scenarios even when all the two-way interactions are included.  The other three tests have power in these two scenarios.  The CrossMST has higher power than the CrossNN test and both have higher power than the crossmatch test.  The conclusion holds for different choices of distances. (The values of $a$ and $b$ were chosen so that these three tests have moderate power to be comparable.)

% The crossmatch test and both proposed tests have power in these two scenarios, and the proposed tests have higher power than the CrossMatch test with CrossMST has the highest power.

% When there are only main effects ($b=0$), the matching on these main effects does a good job and the covariates are well balanced.  We see from the table that all tests reject a few or even no trials.  However, when there are second order effects ($b\neq 0$), only matching the main effects would not make the second order effects balanced and we would hope tests could reject covariate balance.  We see that CrossNN and CrossMatch reject a portion of trials, while neither Hotelling $T^2$ nor Hansen's test reject any of the trials.  In both scenarios that there are second order effects, CrossNN and CrossMST are more powerful than CrossMatch. (The alternatives were chosen so that the tests have moderate power.)

\begin{table}[!htp]
\caption{The proportion of trials (out of 100) that the test rejects covariate balance at the significance level 0.05.} \label{tab:power}
\begin{center}
\begin{tabular}{|c||c|c|c|c|c|c|c|c|c|c|}
\hline
Scenario & $T^2$ & CD & $T^2 \times$ & CD $\times$ & CM$_1$ & NN$_1$ & MST$_1$ & CM$_2$ & NN$_2$ & MST$_2$ \\ \hline \hline
(i) & 0 & 0 & 0.01 & 0.01 & 0.03 & 0.02 & 0.02 & 0.06 & 0.04 & 0.04 \\ \hline
(ii) & 0 & 0 & 0.06 & 0.06 & 0.21 & 0.53 & 0.80 & 0.57 & 0.81 & 0.95 \\ \hline
(iii) & 0 & 0 & 0.02 & 0.02 & 0.22 & 0.56 & 0.79 & 0.50 & 0.81 & 0.97 \\ \hline
(iv) & 0 & 0 & 0.03 & 0.02 & 0.23 & 0.55 & 0.80 & 0.62 & 0.81 & 0.95 \\ \hline  \hline
% 0 & 0.45 & 61 & 0 & 0 & 36 \\ \hline
\end{tabular}
\end{center}
\end{table}

\section{Application to a study of smoking's effect on blood lead}
\label{sec:application}

In \cite{hsu2013calibrating}, the authors studied whether smoking causes an increase in blood lead levels using data from the 2007--2008 U.S. National Health and Nutrition Examination Survey(NHANES) on 679 daily smokers (reported smoking every day for the previous 30 days with an average of 10 more cigarettes per day) and 2661 non-smokers (smoked no cigarettes in previous 30 days and fewer than 100 in lifetime).  We consider this same data and match 679 daily smokers with 679 non-smokers, among the candidate pool of 2661 non-smokers, with the hope that the covariates in the control group are balanced with the exposed/treated group.  We matched using the \texttt{pairmatch()} function in the \texttt{optmatch} package in \texttt{R} \citep{hansen2007flexible} on a distance matrix that combined a caliper on an estimated propensity score with a rank based Mahalanobis distance \citep{rosenbaum2010design}.  Pairs were matched for age, gender, education, income, and race.  Table \ref{tab:smoking} shows means and standardized differences in means (i.e., the difference in means divided by a measure of the average within treatment group standard deviation) before and after matching.

\begin{table}[!htp]
\caption{Summary statistics and covariates balance before and after matching for pairs of a daily smoker and a non-smoker.}\label{tab:smoking}
\begin{center}
\begin{tabular}{lccccc}
\hline  \hline
 & \multicolumn{3}{c}{Before} & \multicolumn{2}{c}{After} \\ \cline{2-4} \cline{5-6}
Covariates & Smoker & Non-smoker  & SD$^*$ & Non-smoker  & SD$^*$ \\
&  ($n=679$) & ($n=2661$) & & ($n=679$) & \\ \hline
Age & 46.71 & 49.78 & -0.188 & 46.79 &  -0.005 \\
Income-to-poverty level & 2.00 & 2.62 & -0.412 & 2.15 & -0.098 \\
\quad Missing,\% & 6.04 & 9.17 & -0.118 & 6.77 & -0.028 \\
Male,\% & 56.85 & 37.81 & \ 0.388 & 56.11 & \ 0.015 \\
Education,\% & \multicolumn{5}{c}{} \\
\quad Less than 9th grade & 11.19 & 13.83 & -0.080 & 9.13 & \ 0.062 \\
\quad 9-11th grade & 26.51 & 13.90 & \ 0.318 & 23.42 & \ 0.078 \\
\quad High school graduate & 32.25 & 23.26 & \ 0.202 & 33.73 & -0.033 \\
\quad Some college & 24.74 & 25.63 & -0.020 & 26.80 & -0.048 \\
\quad College & 5.30 & 23.30 & -0.532 & 6.92 & -0.048 \\
\quad Unknown & 0.00 & 0.08 & -0.039 & 0.00 & \ 0.000 \\
Race,\% & \multicolumn{5}{c}{} \\
\quad White & 65.24 & 40.81 & \ 0.505 & 63.48 & \ 0.037 \\
\quad Black & 19.15 & 19.13 & \ 0.000 & 21.06 & -0.049 \\
\quad Mexican American & 6.48 & 21.35 & -0.440 & 7.07 & -0.017 \\
\quad Other Hispanic & 4.71 & 13.38 & -0.306 & 3.83 & \ 0.031 \\
\quad Other races & 4.42 & 5.34 & -0.043 & 4.57 & -0.007 \\ \hline \hline
\end{tabular}
\end{center}

*SD: Standardized difference $=(\bar{x}_1-\bar{x}_2)/\sqrt{(s_1^2+s_2^2)/2}$, where $\bar{x}_m$ and $s_m^2$ are sample mean and variance for smokers $(m=1)$ and non-smokers $(m=2)$.
\end{table}

We can see from the table that the covariates are quite unbalanced before matching.  For example, some covariates have absolute standardized difference larger than 0.5.  After doing the matching, the covariates seem to be balanced well.  All absolute standardized differences are less than 0.1.  We applied the standard two-sample $t$-test on each of the covariates after matching.  The lowest $p$-value is larger than 0.05.  Hence, the covariates are well balanced separately.

Now the question is whether the covariates are well balanced jointly. We applied the Hotelling $T^2$ test to the matched data and get a $p$-value of 0.69, showing no significant difference.  We also applied the crossmatch test to the matched data and get a $p$-value of 0.014, showing some evidence of a difference between the two groups.  We then applied our proposed methods, CrossNN and CrossMST, to the matched data and get $p$-values of 0.00034 and 0.00077, respectively, which provides very strong evidence rejecting covariate balance.

To see which of these discrepant test results seems to better assess the evidence for covariate imbalance, we explored the data set more.
We regressed the smoking status on the covariates (1) with no interaction term, and (2) with pairwise interactions between covariates.  The categorical variable \texttt{education} has six categories, and none of the treated subjects and matched controls is in the category ``Unknown", so we use four dummy variables to represent \texttt{education} in the model.  The categorical variable \texttt{race} has five categories and is represented by four dummy variables in the model.  This leads to 12 variables in model (1) and 53 additional interaction terms in model (2).

When no interaction term was included in the regression, no coefficient in the regression model is significantly different from 0 except for the intercept.  This is in line with the fact that all covariates are quite well balanced separately.  However, when all pairwise interaction terms are included, quite a few coefficients become significantly different from 0.  They are listed in Table \ref{tab:smoking_interaction}.  We see that, after including the interaction terms, the variables \texttt{age} and \texttt{education} exhibit main effects and there are also extensive interaction effects between \texttt{age} and \texttt{education}.

\begin{table}[!htp]
\caption{Coefficient estimates and standard errors for terms whose coefficients are significantly different 0 ($p$-value less than 0.01).} \label{tab:smoking_interaction}
\begin{center}
\begin{tabular}{lcccc}
\hline \hline
Covariates & Estimate & Standard Error & $t$-value & $p$-value \\ \hline
% \multicolumn{5}{l}{Main effect} \\ \hline
Age & 0.0206 &  0.0063  & 3.268  & 0.00111 \\
Education (Less than 9th grade) &  1.0613 & 0.3791  & 2.799 & 0.00520 \\
Education (9-11th grade) &  0.9355 & 0.3464  & 2.700 & 0.00701 \\
Age $\times$ Education (Less than 9th grade) &  -0.0184 & 0.0047  & -3.924 & $9.17 \times 10^{-5}$ \\
Age $\times$ Education (9-11th grade)  & -0.0189 &  0.0040  & -4.696 & $2.94 \times 10^{-6}$\\
Age $\times$ Education (High school graduate) & -0.0160 & 0.0039 & -4.076 & $4.86\times 10^{-5} $\\
Male $\times$ Education (9-11th grade) & 0.3627 & 0.1406  & 2.580 & 0.00999 \\ \hline \hline
\end{tabular}
\end{center}
\end{table}

Therefore, the covariates of the two groups are not jointly balanced.  Our new tests report very small $p$-values, giving an appropriate alarm whereas the existing tests provided less of an alarm.

\section{Discussion}
\label{sec:conclusion}

We propose two tests, CrossNN and CrossMST, to assess covariate balance in matched observational studies.  Both tests utilize similarity information on the covariates and are powerful for a wide range of alternatives, some of which existing tests have little power for.

Both CrossNN and CrossMST can be extended to denser graphs.  For example, CrossNN can be extended to $k$-NN where each subject finds its first $k$ nearest neighbors.  Similarly, CrossMST can be extended to the $k$-MST, which is the union of $k$ sequentially constructed spanning trees, where the $i$th constructed spanning tree is the spanning tree connecting all observations that minimized the sum of distances across edges subject to the constraint that this spanning tree does not contain any edge in the previously constructed spanning trees (the first constructed spanning tree is just the usual MST).  Choosing $k$ slightly larger than 1 could further boost the power of the tests as the slightly denser graphs contains more similarity information.

%** To be added: Describe what modifications we need to do if the sample sizes of the two groups are different.

In some observational study settings, a control group is assembled by a means other than matching, e.g., choosing certain ranges of the propensity score, but a researcher would still like to know if the assembled control group resembles the treated group, e.g., \cite{sommers2014}.  When the control group and the test group do not have the same number of subjects, we could still use $Z_D$ and $Z_R$ as the test statistics, while the associated expectations, variances, and covariance need to be updated.  The exact formulas are provided in Appendix \ref{sec:unbalanced}.

% Was this added in response to a reviewer?  If not, we could delete to make the discussion more concise.
We have described testing balance among all observed covariates.  Some covariates may be irrelevant to the outcome in the sense described by \cite{pimentel2016} in which case not matching the covariate can diminish the bias from unmeasured covariates.  In such a case, our method could be applied to test balance for the relevant covariates.

% Is there a commonly used distance between curves or a reference worth giving?
Since these tests only need a similarity measure on the sample space, they are not limited to multivariate data, e.g., the covariate could be a curve over time or a network and a distance between these objects could be defined.
% For example, the observation on each covariate does not need to be a value.  It could be a curve over time such as the blood pressure over a month.  As long as a reasonable distance can be defined over the observations, the proposed tests could be applied.

The test statistics may also be extended to do matching, i.e., the test statistics could be used as criteria in finding the `optimal' set of subjects in the control group whose covariates distribution are similar to that of the treatment group.  This could be an interesting direction for future research.

\section*{Acknowledgement}
Hao Chen is supported in part by NSF award DMS-1513653.

 \section*{Supplementary material}
 \label{SM}
Supplementary material available at \Bka\ online includes proofs to Lemma \ref{lemma:EV} and Theorem \ref{thm:asym}.

\appendix

%\appendixone
%\section*{Appendix 1}
%\subsection*{General}

%\appendixtwo
%\section*{Appendix 2}
%\subsection*{Technical details}

\section{Permutation null distribution of the test statistic of CrossMST}
\label{sec:permutation_crossMST}
There are $N-1$ edges in the MST.  Following from \cite{chen2017new}, we have
\allowdisplaybreaks
\begin{align}
& \bE(R_1) = \bE(R_2) = \tfrac{N-2}{4}, \\
& \bV(R_1) = \bV(R_2) =  \tfrac{1}{16(N-3)}\left(-(N-2)(N-6) + 2C_3 \tfrac{N(N-4)}{N-1} \right),  \\
& \bCov(R_1,R_2) = \tfrac{N-2}{16(N-3)}\left(3(N-2)-2C_3 \tfrac{N}{N-1}\right),
\end{align}
where $C_3$ is the number of edge pairs that share a common node.

Let $Z_{R,1} = \frac{R_1-\bE(R_1)}{\sqrt{\bV(R_1)}}, \quad Z_{R,2} = \frac{R_2-\bE(R_2)}{\sqrt{\bV(R_2)}}.$
Then the test statistic $Z_R = \max(Z_{R,1},Z_{R,2})$ is equivalent to $R_M$.

Let $G_{MST}$ be the set of edges in the MST.
\begin{align}
  A_{e,MST} & = \{e\} \cup \{e^\prime\in G_{MST}: e^\prime \text{ and } e \text{ share a node}\},\nonumber \\
  B_{e,MST} & = A_e \cup \{e^{\prime\prime}\in G_{MST}: \exists \  e^\prime\in A_{e,MST}, \text{ such that } e^{\prime\prime} \text{ and } e^\prime \text{ share a node}\}.\nonumber
\end{align}
Then $A_e$ is the subgraph in $G$ that connects to edge $e$, and $B_e$ is the subgraph in $G$ that connects to any edge in $A_e$.

Then, following from \cite{chen2017new}, we have that, if $\sum_{e\in G}|A_{e,MST}||B_{e,MST}|=o(N^{1.5})$, $C_3-N=O(N)$, then under permutation null distribution, as $N\rightarrow\infty$, $(Z_{R,1},Z_{R,2})^\prime$
converges to a bivariate Gaussian distribution with mean $\mathbf{0}$ and covariance matrix $\left[\begin{array}{cc} 1 & \rho_R \\ \rho_R & 1 \end{array}\right]$, where $\rho_R = (3-2q_3)/(2q_3-1)$ with $q_3 \overset{\Delta}{=} \lim_{N\rightarrow} C_3/N$.

We then can easily calculate the asymptotic $p$-value of $Z_R$.  That is,
\begin{align*}
P(Z_R\geq z) & = 1-P(Z_R<z) = 1-P(Z_{R,1}<z, Z_{R,2}<z),
\end{align*}
and $P(Z_{R,1}<z, Z_{R,2}<z)$ is calculated using function \texttt{pmvnorm()} in \texttt{R} package \texttt{mvtnorm}.  In practice, we use the finite sample version of the covariance between $Z_{R,1}$ and $Z_{R,2}$, that is, we use $$\rho_{R,N}=\frac{3(N-2)-2C_3 N/(N-1)}{-(N-6)+2C_3 N(N-4)/(N-1)/(N-2)}$$ rather than $\rho$ when calculating $P(Z_{R,1}<z, Z_{R,2}<z)$.  It is easy to see that $\lim_{N\rightarrow\infty}\rho_{R,N} = \rho$.

% \subsubsection{Accuracy of asymptotic results on finite samples}
% \label{sec:accuracy}

\begin{figure}[!htp]
\begin{center}
\includegraphics[width=.40\textwidth]{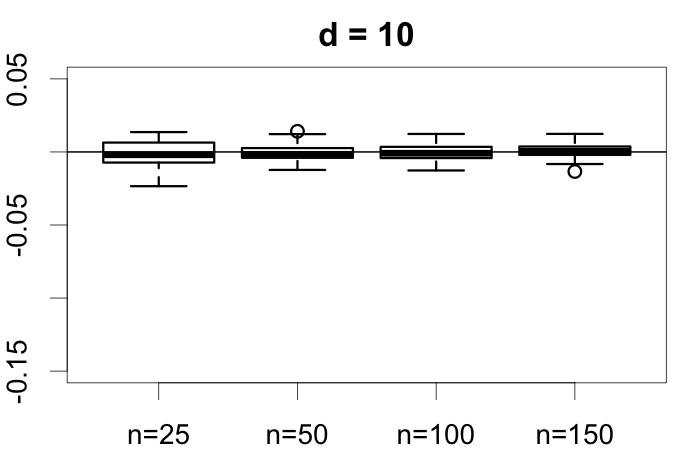} \quad
\includegraphics[width=.40\textwidth]{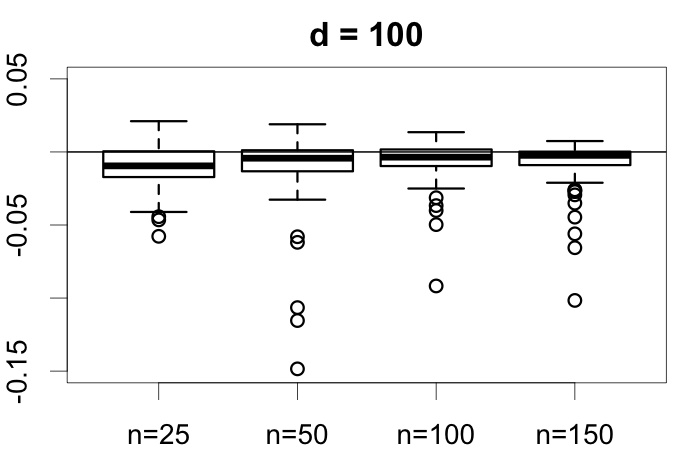}
\end{center}
\caption{Boxplots of the differences between the asymptotic $p$-value and the $p$-value calculated from 10,000 permutations (asymptotic $p$-value minus permutation $p$-value).}\label{fig:pvalue_R}
\end{figure}

We check how the asymptotic $p$-value approximation works for finite sample sizes.  We compare the asymptotic $p$-value to $p$-value obtained through 10,000 permutation directly.  The boxplots of the differences under different sample sizes are shown in Figure \ref{fig:pvalue_R}.  Similarly to CrossNN, the asymptotic $p$-value approximation works very well for small sample sizes when the dimension is low ($d=10$).  The approximation becomes worse when the dimension becomes high ($d=100$), while it is reasonably accurate for sample sizes in hundreds.

\section{Statistics for unbalanced sample sizes}
\label{sec:unbalanced}
When the control group and the treatment group have different numbers of subjects, we could still use $Z_D$ and $Z_R$ as the test statistics, while the associated expectations, variances, and covariances need to be updated.  They can be derived similarly as the equal sample size case shown in supplemental materials, and their exact analytic formulas are provided below.  In the following, $n_1$ is the number of subjects in the treatment group, and $n_2$ is the number of subjects in the control group.

\allowdisplaybreaks
\subsection{CrossNN}
In the default CrossNN approach where each observation finds its nearest neighbor, we have,
\begin{align*}
& \bE(D_{11}) = \tfrac{n_1(n_1-1)}{N-1}, \quad \bE(D_{22}) = \tfrac{n_2(n_2-1)}{N-1}, \\
& \bV(D_{11}) = \tfrac{n_1(n_1-1)n_2(n_2-1)}{N(N-1)(N-2)(N-3)}\left( \tfrac{N(N-3)}{N-1} + 2C_1 + 2C_2  \tfrac{n_1-2}{n_2-1} \right), \\
& \bV(D_{22}) = \tfrac{n_1(n_1-1)n_2(n_2-1)}{N(N-1)(N-2)(N-3)}\left( \tfrac{N(N-3)}{N-1} + 2C_1 + 2C_2 \tfrac{n_2-2}{n_1-1} \right), \\
& \bCov(D_{11}, D_{22}) = \tfrac{n_1(n_1-1)n_2(n_2-1)}{N(N-1)(N-2)(N-3)}\left(  \tfrac{N(N-3)}{N-1} + 2C_1 - 2C_2 \right).
\end{align*}

If $k$-NN is used, we could further extend the test by defining $a_{ij}$ to be 1 if $\by_j$ is among the first $k$ NNs of $\by_i$ and 0 otherwise.  Then, $D_{11}$ is the number of edges in the $k$-NN graph (each observation points to its first $k$ nearest neighbors) such that both the starting and ending points of the edge are from the treatment group, and $D_{22}$ is that for the control group.  The more general formulas of the expectations, variances, and covariance are provided below.  In the following, $C_1$ and $C_2$ are also defined in the extended sense that $C_1$ is the pair of observations points to each other in the $k$-NN graph and $C_2$ is the pair of edges in the $k$-NN graph such that they point to the same observation.

\begin{align*}
& \bE(D_{11}) = \tfrac{k n_1(n_1-1)}{N-1}, \\
& \bE(D_{11}) = \tfrac{k n_2(n_2-1)}{N-1}, \\
& \bV(D_{11}) = \tfrac{n_1(n_1-1)n_2(n_2-1)}{N(N-1)(N-2)(N-3)}\left( kN + 2C_1 + \tfrac{n_1-2}{n_2-1}(2C_2 + kN -k^2 N) - \tfrac{2}{N-1}  k^2 N \right), \\
& \bV(D_{22}) = \tfrac{n_1(n_1-1)n_2(n_2-1)}{N(N-1)(N-2)(N-3)}\left( kN + 2C_1 + \tfrac{n_2-2}{n_1-1}(2C_2 + kN -k^2 N) - \tfrac{2}{N-1}  k^2 N \right), \\
& \bCov(D_{11}, D_{22}) = \tfrac{n_1(n_1-1)n_2(n_2-1)}{N(N-1)(N-2)(N-3)}\left(  2C_1 - 2C_2 + \tfrac{k^2 N(N-3)}{N-1} \right).
\end{align*}

\subsection{CrossMST}
Here, we provide the general formulas for the expectations, variances, and covariance in the CrossMST approach when $k$-MST is used and $n_1$ and $n_2$ could be different.  In the following, $C_3$ is similarly defined as in Appendix \ref{sec:permutation_crossMST} that it is the number of edge pairs in $k$-MST that share a common node.

\begin{align*}
& \bE(R_1) = \tfrac{k n_1(n_1-1)}{N}, \\
& \bE(R_2) = \tfrac{k n_2(n_2-1)}{N}, \\
& \bV(R_1) = \tfrac{n_1(n_1-1)n_2(n_2-1)}{N(N-1)(N-2)(N-3)}\left( \tfrac{n_1-2}{n_2-1}\left(2C_3 + 2k(N-1) - \tfrac{4k^2(N-1)^2}{N} \right)  + \tfrac{k(N-1)(N-2k)}{N} \right), \\
& \bV(R_2) = \tfrac{n_1(n_1-1)n_2(n_2-1)}{N(N-1)(N-2)(N-3)}\left( \tfrac{n_2-2}{n_1-1}\left(2C_3 + 2k(N-1) - \tfrac{4k^2(N-1)^2}{N} \right) + \tfrac{k(N-1)(N-2k)}{N} \right), \\
& \bCov(R_1, R_2) = \tfrac{n_1(n_1-1)n_2(n_2-1)}{N(N-1)(N-2)(N-3)}\left( - 2C_3 + \tfrac{k(N-1)(4kN-N-6k)}{N} \right).
\end{align*}
%
% to accommodate this.  Let $n_1$ be the number of cases and $n_2$ be the number controls.  Suppose $n_2 = c n_1$, then the statistic based on $k$-NN could be modified as
%$$\max(c D_{11}, D_{22}) $$
%and the statistic based on $k$-MST could be modified as
%$$\max(c R_1, R_2)$$
%
%** Another possible choice for $k$-NN: $\max\left(\left| \frac{D_{11}-D_{22}-(\bE(D_{11})-\bE(D_{22}))}{\sqrt{\bV(D_{11})+\bV(D_{22})-2\textbf{Cov}(D_{11}, D_{22})}} \right|, \frac{cD_{11}+D_{22}-(c\bE(D_{11})+\bE(D_{22}))}{\sqrt{c^2\bV(D_{11})+\bV(D_{22})+2c\textbf{Cov}(D_{11}, D_{22})}}
%\right)$ or a more flexible version $\max\left(\left| \frac{D_{11}-D_{22}-(\bE(D_{11})-\bE(D_{22}))}{\sqrt{\bV(D_{11})+\bV(D_{22})-2\textbf{Cov}(D_{11}, D_{22})}} \right|, \kappa \frac{cD_{11}+D_{22}-(c\bE(D_{11})+\bE(D_{22}))}{\sqrt{c^2\bV(D_{11})+\bV(D_{22})+2c\textbf{Cov}(D_{11}, D_{22})}}
%\right)$
%

%Another possible choice for $k$-MST: $\max(|R_1-R_2|, cR_1 + R_2)$ or a more flexible version  $\max(|R_1-R_2|, \kappa(cR_1 + R_2))$.

\bibliographystyle{biometrika}
\bibliography{balanceCheck}

\end{document}